\documentclass[conference]{IEEEtran}
\usepackage{tabularx}
\usepackage{listings}
\usepackage[table]{xcolor}
\usepackage{cite}
\usepackage{graphicx}
\DeclareGraphicsExtensions{.eps}
\usepackage{array}
\usepackage[caption=false,font=footnotesize]{subfig}
\usepackage{fixltx2e}
\usepackage{url}
\usepackage[nolist]{acronym}

\newcommand{\epsep}{\rceil}
\newcommand{\olsep}{\|}
\newcommand{\nolsep}{|}
\newcommand{\ecmspace}{\,}
\newcommand{\ecm}[6]{\mbox{$\left\{{#1}\ecmspace\olsep\ecmspace {#2}\ecmspace\nolsep\ecmspace {#3}\ecmspace\nolsep\ecmspace {#4}\ecmspace\nolsep\ecmspace {#5}\right\}\ecmspace{#6}$}}
\newcommand{\ecmp}[5]{\mbox{$\left\{{#1}\ecmspace\epsep\ecmspace {#2}\ecmspace\epsep\ecmspace {#3}\ecmspace\epsep\ecmspace {#4}\right\}\ecmspace{#5}$}}

\begin{document}

\begin{acronym}[CISC]
    \acro{AGU}{Address Generation Unit}
    \acro{AVX}{Advanced Vector Extensions}
    \acro{BPU}{Branch Prediction Unit}
    \acro{CISC}{Complex Instruction Set Computer}
    \acro{CL}{Cache Line}
    \acro{CoD}{Cluster on Die}
    \acro{DP}{Double Precision}
    \acro{ECM}{Execution-Cache-Memory}
    \acro{EDP}{Energy-Delay Product}
    \acro{FIVR}{Fully Integrated Voltage Regulators}
    \acro{FMA}{Fused Multiply-Add}
    \acro{Flops}{Floating-Point Operations}
    \acro{HA}{Home Agent}
    \acro{ILP}{Instruction Level Parallelism}
    \acro{IMCI}{Initial Many Core Instructions}
    \acro{ISA}{Instruction Set Architecture}
    \acro{LFB}{Line Fill Buffer}
    \acro{LLC}{Last-Level Cache}
    \acro{MC}{Memory Controller}
    \acro{NUMA}{Non-Uniform Memory Access}
    \acro{OoO}{Out-of-Order}
    \acro{PCIe}{Peripheral Component Interconnect Express}
    \acro{PRF}{Physical Register File}
    \acro{QPI}{QuickPath Interconnect}
    \acro{RFO}{Read for Ownership}
    \acro{RISC}{Reduced Instruction Set Computer}
    \acro{SIMD}{Single Instruction Multiple Data}
    \acro{SP}{Single Precision}
    \acro{SSE}{Streaming SIMD Extensions}
    \acro{UFS}{Uncore Frequency Scaling}
\end{acronym}

\title{Execution-Cache-Memory Performance Model: Introduction and Validation}
\author{\IEEEauthorblockN{Johannes Hofmann}
\IEEEauthorblockA{Chair for Computer Architecture\\
University Erlangen--Nuremberg\\
Email: johannes.hofmann@fau.de}
\and
\IEEEauthorblockN{Jan Eitzinger}
\IEEEauthorblockA{Erlangen Regional Computing Center (RRZE)\\
University Erlangen--Nuremberg\\
Email: jan.eitzinger@fau.de}
\and
\IEEEauthorblockN{Dietmar Fey}
\IEEEauthorblockA{Chair for Computer Architecture\\
University Erlangen--Nuremberg\\
Email: dietmar.fey@fau.de}
}
\maketitle
\begin{abstract}
    This report serves two purposes: To introduce and validate the \ac{ECM}
    performance model and to provide a thorough analysis of current Intel
    processor architectures with a special emphasis on Intel Xeon Haswell-EP.
    The \ac{ECM} model is a simple analytical performance model which focuses on
    basic architectural resources. The architectural analysis and model
    predictions are showcased and validated using a set of elementary
    microbenchmarks.
\end{abstract}

\section{Introduction}
Today's processor architectures appear complex and in-transparent to software
developers. While the machine alone is already complicated the major complexity
is introduced by the interaction between software and hardware. Processor
architectures are in principle still based on the stored-program computer
design. In this design the actions of the computer are controlled by a
sequential stream of instructions.  The instructions are plain numbers and
therefore equivalent to data.  Consequently they can be stored in the same
memory. This is a strictly sequential design and the main focus of a system
designer is to increase the instruction throughput rate. Modern processors
nevertheless employ parallelism on various levels to increase throughput:
instruction level, data processing level, and task level.  It is possible to
formulate a very simple performance model which reduces the processor to its
elementary resources: instruction execution and data transfers. To set up the
model one needs to determine the time it takes to execute a given instruction
sequence on a given processor core and to transfer the data, which is required
to do so. Intimate knowledge about processor architecture, cache and memory
architectures are necessary to do this. It is still a worthwhile effort as the
model gives in-depth insights about bottlenecks, runtime contributions, and
optimization opportunities of a code on a specific processor. This report
introduces the model and provides a thorough analysis Intel's latest Haswell-EP
processor architectures.

\section{Processor Microarchitecture}
Intel processors use the Intel~64\footnote{Intel~64 is Intel's implementation
of x86-64, the 64\,bit version of the x86 instruction set. x86-64 was
originally developed by AMD under the AMD64 moniker and while Intel 64 and
AMD64 are almost identical, there exist minor differences that warrant
differentiation.} \ac{ISA} which is a so called \ac{CISC} architecture.  It
originates from the late 70s and initially was a 16\,bit \ac{ISA}. During its
long history it was extended to 32\,bit and finally 64\,bit. The \ac{ISA}
contains complex instructions that can have variable number of operands, have
variable opcode length and allow for address references in almost all
instructions.  To execute this type of \ac{ISA} on a modern processor with
aggressive \ac{ILP} the instructions must be converted on the fly into a
\ac{RISC} like internal instruction representation. Intel refers to these
instructions as micro-operations, $\mu$ops for short. Fortunately decoding of
\ac{CISC} instructions to $\mu$ops works so well that it does not negatively
impact instruction throughput.  Please note that when talking about
instructions we mean the RISC like internal instructions called $\mu$ops and
not Intel 64 ISA instructions. In the following, we will describe the most
important techniques to increase performance in contemporary processor
architectures.

{\bf ILP} - As any modern processor, Intel processors aggressively employ
parallel instruction execution within the strictly sequential instruction
stream. This parallelism is exploited dynamically by hardware during execution
and requires no programmer or compiler intervention. \ac{ILP} comes in two
flavors: Pipelining and superscalar execution. Pipelining executes different
stages of multiple instructions simultaneously. In superscalar designs multiple
execution pipelines exist and can be active and execute instructions at the
same time.  Where pipelining enables an instruction throughput of one per cycle
superscalar execution allows to retire multiple instructions per cycle. Due to
dependencies between instructions the degree of ILP that can be leveraged
heavily depends on the instruction mix of a particular code and is typically
limited.
In order to exploit even more parallelism most modern general purpose
processors support \ac{OoO} execution.  In \ac{OoO} execution the processor may
change the order in which instructions are executed as long as semantic
equivalency to the original ordering is guaranteed.  Common codes involve many
conditional branches which severely limit the size of the instruction window to
apply \ac{ILP} to. Therefore \ac{OoO} execution is usually combined with
speculative execution. This technique attempts to predict the outcome of
branches and speculatively executes the forecast code path before the outcome
is known. This may involve executing unnecessary instructions but enables to
exploit ILP across branches, which is crucial for loop bodies of limited size.
\ac{ILP} is still a major technology for generating high performance, but it is
not a dominating driver of performance improvements anymore.  Implementations
already are highly optimized and in all but a selected special cases work very
well.

{\bf SIMD} - Another level of parallelism are data parallel instructions which
simultaneously perform the same operation on multiple data items. To make use
of this architectural feature, dedicated so called \ac{SIMD} instructions have
to be used by the software. Those \ac{SIMD} instructions are provided by means
of instruction set extensions to the core \ac{ISA}. Currently \ac{SIMD} is a
major driver for performance. The reason is that it is relatively simple to
implement in hardware since the overall instruction throughput is not altered.
\ac{SIMD} is characterized by its register width. The current width is 256\,bit
(\ac{AVX}) with 512\,bit already employed in Intel's Knights Corner
architecture (\ac{IMCI}) and announced for regular Xeon processors with the
advent of Skylake (AVX-512).  Apart from performing multiple operations in a
single instruction another benefit of \ac{SIMD} is that of loading and storing
data block-wise. The same amount of work can be done with a factor less
instructions. It can be already predicted that the role of \ac{SIMD} as a major
driver for performance comes to an end with the introduction of 512\,bit SIMD
width.

{\bf Multicore chips} - Moore's law continues to drive the number of
transistors which can be packed on a single chip. During the 90s the increase
in register count enabled by the shrinking of manufacturing size was
accompanied by increasing clock speed as a major way to increase the
performance of processors.  In the early 2000s a paradigm shift occurred. The
vendors did not manage to further increase clock speed without running into
cooling issues. The answer to this dilemma was to put multiple (processor)
cores on the same die.  Early designs had completely separated caches and only
shared main memory access. Later some of the caches were private and some
shared. For a programmer a multicore processor feels like a multi-processor SMP
system. Parallel programming is required to leverage the performance. The core
is now a building block and a major engineering effort is put into how to
interconnect cores on the die and how to route data from main memory
controllers to the caches.  At the moment a still moderate number of cores is
put on one die connected by one or more segmented ring buses. The \ac{LLC} is
usually also segmented.  Multiple memory controllers with multiple channels are
connected to the bus to inject data.  Already now and even more in the future
the system on a chip designs will be the performance defining feature of a
processor.  On Intel chips the cores including caches private to a core are
logically separated from shared entities on the chip. Those shared entities are
grouped in the so called uncore. \ac{LLC}-segments, ring-bus, on-board
interconnects and memory controllers are all part of the uncore.

{\bf System Design} - A compute node employs elementary building blocks on
different levels. A core is built of multiple executions units, multiple cores
form a die, there might be multiple dies on one package (socket), and finally a
node might contain multiple sockets. The trend of the system on a chip designs
transfers more and more components which where formerly offered in the
Northbridge on the motherboard or by separate chips onto the processor die.
This involves not only the memory controllers but also \ac{PCIe} interfaces, network
interfaces, and GPUs. For a programmer this adds additional complexity. For
memory access data locality becomes an issue as main memory is distributed in
multiple locality domains (ccNUMA). IO and network access performance might
depend on the origin of the request within the system.

The central part of a microarchitecture are its scheduler and execution units.
With the introduction of the Pentium Pro in 1995 Intel provided a very
accessible abstraction for the scheduler. The scheduler can issue instructions
to so called ports. There is a maximum number of instructions the scheduler can
issue in a single cycle. Behind every port there can be one or more execution
units. The maximum number of instructions which can retire may be different
from the number of instructions which can be issued. Because of speculative
execution it makes sense to issue more instruction than can retire.  This
layout allows an analytical access to predict the instruction throughput of a
given instruction sequence assuming that there are no hazards and dependencies
among instructions.

Changes in microarchitectures can be grouped in incremental, capability and
functional changes. An incremental change is e.g. to add more entries to a
queue the benefit usually is in the single digit percentage range. Capability
changes are e.g. increasing SIMD width, adding more execution units or widen a
data path. Benefits range from percentage improvements to factors.  Functional
changes are adding new instructions introducing a new functionality, e.g.
gather/scatter of \ac{FMA} instructions. In recent years with energy
consumption a new dimension was added in microarchitecture design. This is
driven on one side by the rise of mobile devises where energy consumption is a
primary requirement for processors but also in Supercomputing with energy
consumption limiting the economic feasibility of large scale machines.

\section{Haswell Microarchitecture}

\subsection{Core Pipeline}

\begin{figure}[tb]
\includegraphics[width=\linewidth]{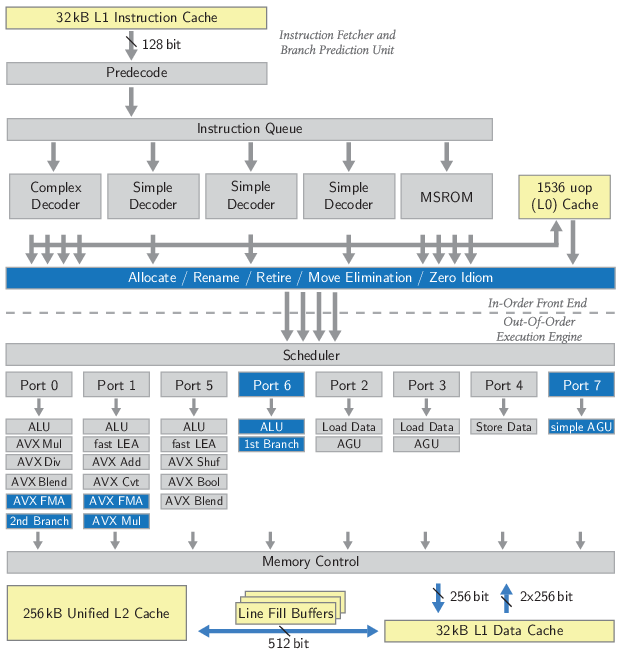}
\caption{Core layout for the Haswell Microarchitecture}
\label{fig:hsw-pipeline}
\end{figure}

Figure \ref{fig:hsw-pipeline} illustrates the simplified core layout of the
Haswell microarchitecture. As all modern designs, this microarchitecture uses a
Harvard design for the innermost cache level, i.e. instructions and data are
stored in separate caches. Starting with the L2 cache it is based on a von
Neumann design with unified caches for instructions and data.  The core fetches
instructions in chunks of 16\,byte from the address provided by the
\ac{BPU}---typically this address is just a 16\,byte increment of the last
address from which data was fetched; in the case of branches it will be the
address of instructions that are the most likely to be executed. After
instructions have been fetched, a pre-decoder determines the bounds of the
various instructions that were included in a given 16\,byte block. In the next
phase, decoding from \ac{CISC} instructions to $\mu$ops occurs. A simple
example would be a single arithmetic operation with memory address as operand
(e.g. \texttt{vaddpd ymm0, ymm0, [rax+r11*8]}) that is split into two $\mu$ops:
one dedicated load operation and a dedicated arithmetic operation with
register-only operands.  This decoding phase is superscalar, with one complex
and three simple complex decoders; also featured is a MSROM decoder which is
responsible for seldom used \ac{RISC} instructions that decode to more that
4~$\mu$ops.  Decoded $\mu$ops are stored in the $\mu$op cache, which can hold
up to 1536 micro-ops, and enables the reuse of previously decoded instructions,
e.g. in the event of loops. The motivation for this cache is energy saving:
whenever micro-ops from the cache are used, the legacy decode pipeline can be
powered down.

Before $\mu$ops leave the in-order front-end, the renamer allocates resources
from the \ac{PRF} to each instruction. One of the improvements of Haswell in
this phase is the elimination of register-registers moves through register
renaming without having to issue any $\mu$ops.  Dependency breaking idioms such
as zero idioms (e.g. \texttt{vxorpd}) and the ones idiom (\texttt{cmpeq}) can
improve instruction parallelism by eliminating false dependencies: The
renamer notices whenever an architectural registers (e.g. \texttt{ymm0}) is set
to zero and will assign a fresh register from the \ac{PRF} to it; the \ac{OoO}
scheduler will thus never see a false dependency. The size of the \ac{OoO}
window has been increased from 168 to 192 micro-ops in Haswell.

The width of all three data paths between the L1 cache and processor registers
has been doubled in size from 16\,B to 32\,B.  This means that AVX loads and
stores (32\,B in size) can now retire in a single clock cycle as opposed to two
clock cycles required on the Sandy and Ivy Bridge architectures.  The data path
between the L1 and L2 caches has also seen a doubling in size---at least on for
transfers from L2 to the L1 cache; our measurements indicate that evictions
still occur at a bandwidth of 32\,B/c.

While the core is still limited to retiring only four $\mu$ops per cycle, the
number of ports has been increased from six to eight in Haswell (shown in blue
in Fig.~\ref{fig:hsw-pipeline}).  The newly introduced port~6 contains the
primary branch unit; a secondary unit has been added to port~0. In previous
designs only a single branch unit was available and located on port~5.  By
moving it to a dedicated port in the new design, port~5---which is the only
port that can perform AVX shuffle operations---is freed up. Adding a secondary
branch units benefits branch-intensive codes.  The other new port is port~7,
which houses a so-called simple \ac{AGU}. This unit was made necessary by the
increase in register-L1 bandwidth. Using AVX on Sandy Bridge and Ivy Bridge,
two \ac{AGU}s were sufficient, because each load or store required two cycles
to complete, not making it necessary to compute three new addresses every
cycle, but only every second cycle. With Haswell this has changed, because
potentially a maximum of three load/store operations can now retire in a single
cycle, making a third \ac{AGU} necessary. Unfortunately, this simple \ac{AGU}
can not perform the necessary addressing operations required for streaming
kernels on its own (see Section~\ref{sec:res:triads} for more details).

Apart from adding additional ports, Intel also extended existing ports with new
functionality.  Operations introduced by the \ac{FMA} \ac{ISA} extension are
handled by two new, AVX-capable units on ports~0 and~1. Haswell is also the
first architecture to feature the AVX2 ISA extension. Because AVX introduced
256\,bit SIMD operations only for \ac{SP} and \ac{DP} floating-point data
types, AVX2 extends the set of 256\,bit SIMD operations to several integer data
types.  Haswell also saw the introduction of a second \ac{AVX} multiplication
unit on port~1. 

\subsection{Package Layout}
\label{sec:hsw:package}

\begin{figure}[tb]
\includegraphics[width=\linewidth]{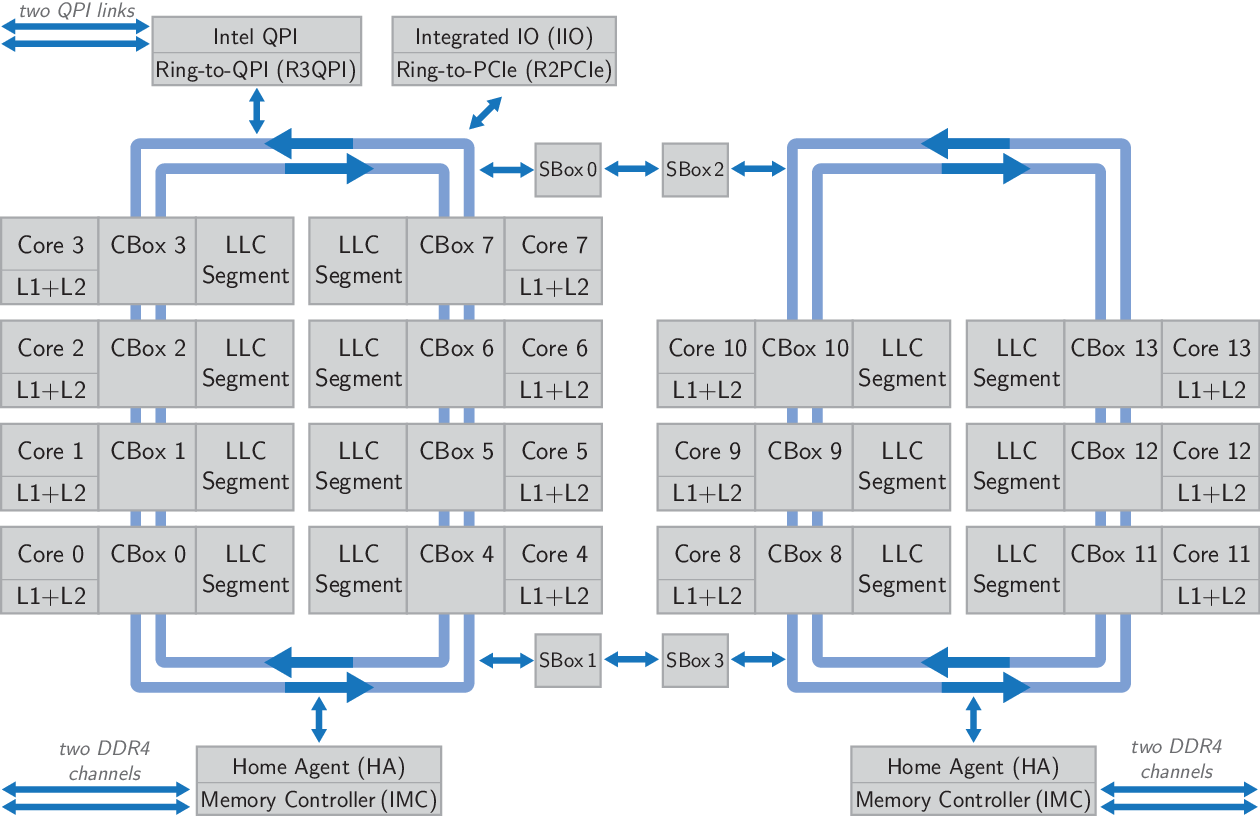}
\caption{Chip layout for the Haswell-EP Microarchitecture}
\label{fig:hsw-package}
\end{figure}

Figure~\ref{fig:hsw-package} shows the layout of a 14-core Haswell processor
package. Apart from the processor cores, the package consists of what Intel
refers to as the uncore. Attached to each core and its private L1 and L2
caches, we find a \ac{LLC} segment, that can hold 2.5\,MB of data. This
physical proximity of core and cache segment does however not imply that data
used by a core is stored exclusively or even preferably in its LLC segment.
Data is placed in all LLC segments according to a proprietary hash function
that is supposed provide uniform distribution of data and prevent hotspots for
a wide range of data access patterns. An added benefit of this design is that
single-threaded applications can make use of all available LLC.

The cores and \ac{LLC} segments are connected to a bidirectional ring
interconnect that can transfer one \ac{CL} (64\,B in size) every two cycles in
each direction.  In order to reduce latency, the cores are arranged to form two
rings, which are connected via two queues. Each ring has associated to it a
\ac{HA} which is responsible for cache snooping operations and reordering of
memory requests to optimize memory performance. Attached to each \ac{HA} is a
\ac{MC}, each featuring two 8\,byte-wide DDR4 memory channels.  Also accessible
via the ring interconnect are the one-die \ac{PCIe} and \ac{QPI} facilities.

Haswell also saw the introduction of an on-die \ac{FIVR}.  This \ac{FIVR}
allows Haswell to draw significantly less power than the previous Ivy Bridge
microarchitecture, because it allows for faster entering and exiting
power-saving states. It also allows a more fine-grained control of CPU states:
instead of globally setting the CPU frequencies for all cores on a package,
Haswell can now set cores frequencies and sleep states individually.

\subsection{Uncore Frequency Scaling}

When Intel first introduced the shared on-die \ac{LLC} with Nehalem, it
maintained distinct clock domains for CPU cores and the uncore, which houses
the \ac{LLC}, because this cache cache was not considered latency sensitive and
could thus run at a lower frequency thereby saving power. In the next
microarchitecture, Sandy Bridge, Intel changed this design and made the uncore
run at the same clock frequency as the CPU cores. While this drastically
benefited latency, it brought with it the problem of on-die graphics accessing
data from the \ac{LLC} with low performance when CPU cores were in power saving
mode and clocked the uncore down along with them. Ivy Bridge tried to solve
this problem with a dedicated L3 graphics cache, but eventually data would have
to be brought in from the regular L3 cache. In the new Haswell
microarchitecture, Intel moved back to the Nehalem design: having two separate
clock domains for core and uncore.  Haswell also offers a feature called
\ac{UFS}, in which the uncore frequency is dynamically scaled based on the
number of stall cycles in the CPU cores.  Although reintroducing high
latencies, the separate clock domain for the uncore offers a significant
potential for power saving, especially for serial codes.

\begin{figure}[tb]
\includegraphics[width=\linewidth]{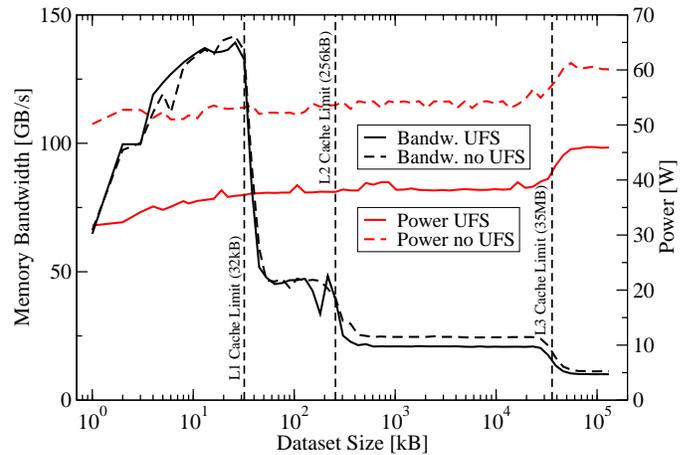}
\caption{Impact of \ac{UFS} on Bandwidth and Power Usage.}
\label{fig:hsw-ufs}
\end{figure}

Figure~\ref{fig:hsw-ufs} shows the sustained bandwidth (left $y$-axis) measured
for the Sch\"onauer vector triad (cf. Table~\ref{tab:benchmarks}) using a
single core along with the power consumption (right $y$-axis) for varying
dataset sizes. As expected the performance is not influenced by whether UFS is
active or not when data resides in a core's private caches (L1+L2). Although we
observe a difference in performance as soon as the LLC is involved, the
performance impact is very limited. While the bandwidth drops from 24 to
21\,GB/s (about 13\%) in the LLC, power usage is reduced from 55 to 40\,W
(about 27\%). In multicore scenarios that work on data in the LLC or main
memory this effect can no longer be observed because the uncore is dynamically
adjusted to run at the maximum clock speed of 3\,GHz in order to satisfy demand
from all cores.

\subsection{Memory}

\begin{figure}[htb]
\includegraphics[width=\linewidth]{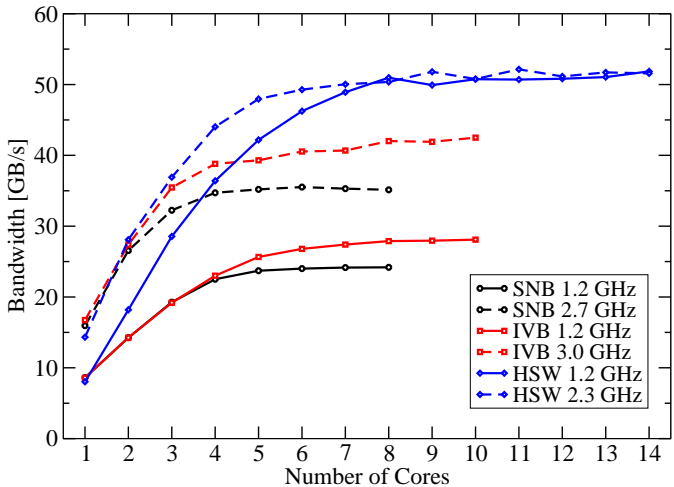}
\caption{Stream Triad Bandwidth as Function of Frequency.}
\label{fig:bw-freq-comparison-stream-triad}
\end{figure}

Microarchitectures preceding Haswell show a strong correlation between CPU
frequency and the achievable sustained memory bandwidth.  This behaviour is
demonstrated in Figure~\ref{fig:bw-freq-comparison-stream-triad}, which shows
the measured chip bandwidth for the Stream Triad---adjusted by a factor of 1.3
to account for write-allocates---on the Sandy Bridge, Ivy Bridge, and Haswell
microarchitectures.

\begin{figure*}[tb]
\includegraphics[width=\linewidth]{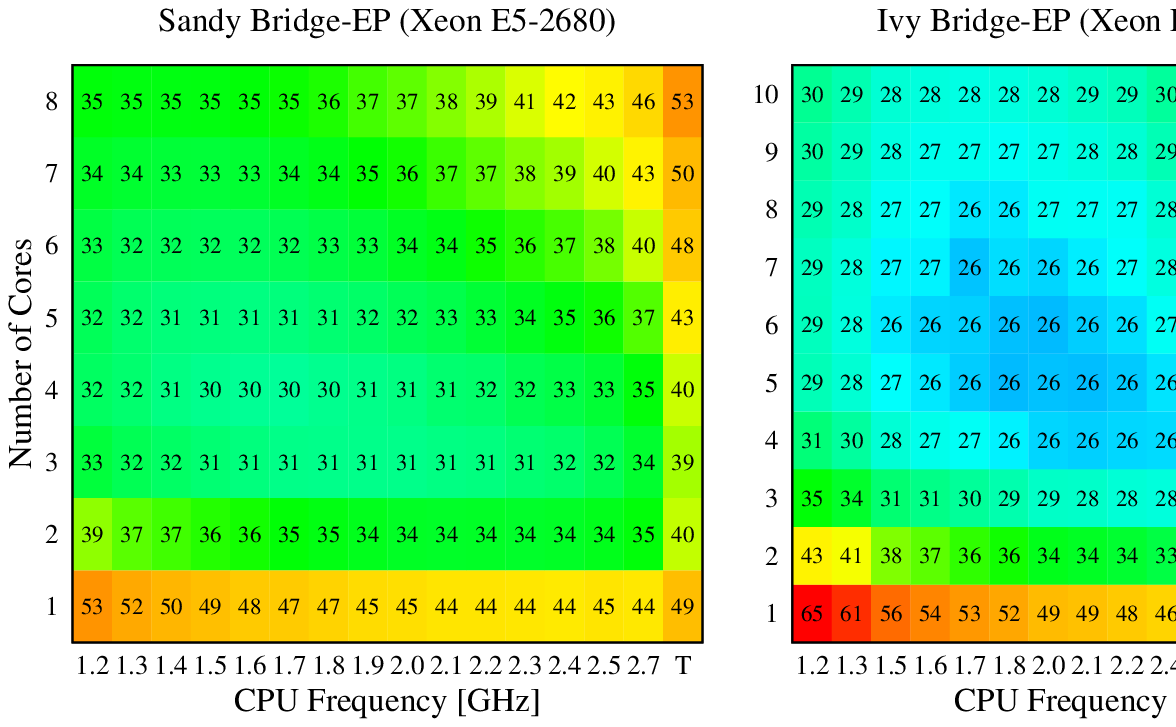}
\caption{Energy-to-Solution for Stream Triad (10\,GB dataset size) on Selected Microarchitectures.}
\label{fig:heat-e2s}
\end{figure*}

\begin{figure*}[tb]
\includegraphics[width=\linewidth]{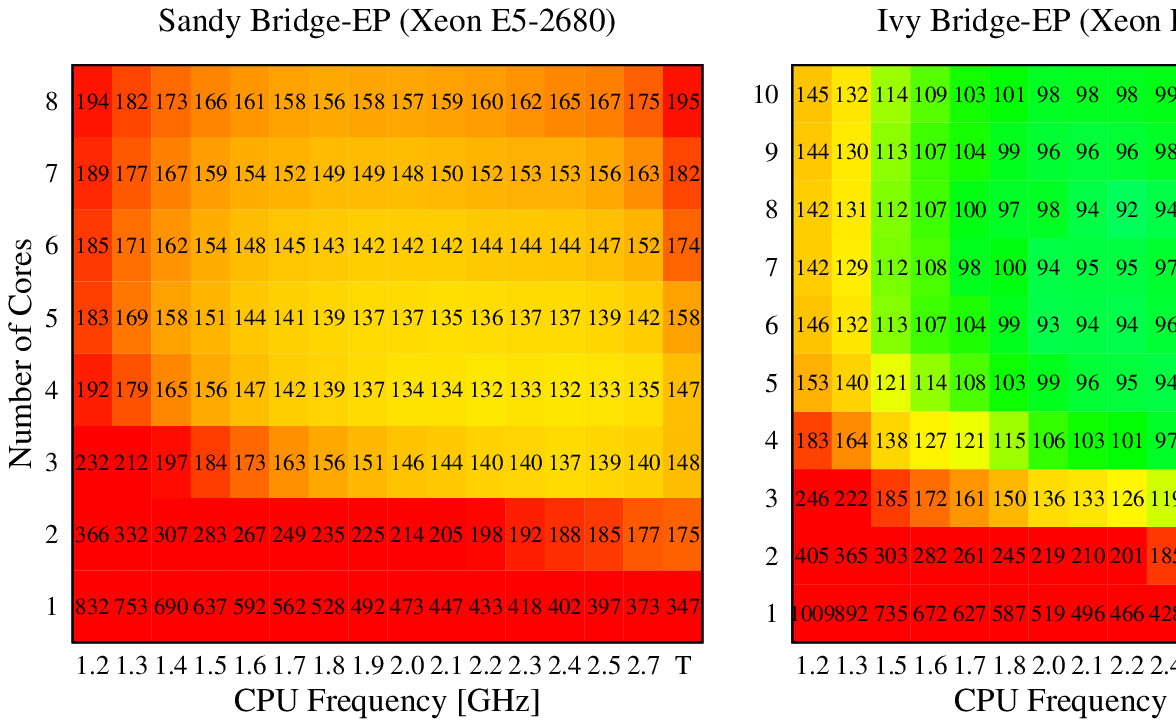}
\caption{\ac{EDP} for Stream Triad (10\,GB dataset size) on Selected Microarchitectures.}
\label{fig:heat-edp}
\end{figure*}

For each system, the bandwidth was measured using the lowest possible frequency
(1.2\,GHz in each case) and the advertised nominal clock speed.  While Sandy
Bridge can achieve a sustained bandwidth of 35.5\,GB/s when clocked at
2.7\,GHz, the result using 1.2\,GHz is only 24.2\,GHz---just $2/3$ of the
best-case chip bandwidth! On Ivy Bridge, the nominal clock speed of 3.0\,GHz
delivers a sustained chip bandwidth of 42.5\,GHz; at 1.2\,GHz the performance
degrades to 28.1\,GB/s---again, just $2/3$ of the best-case chip bandwidth.
For Haswell, we observe that the sustained bandwidth of 52.3\,GHz is identical
in both cases. Even the saturation point---the number of cores required to
reach the sustained socket bandwidth---of 7--8 cores is almost identical.
Bearing in mind that CPU frequency is the most important variable influencing
package power usage, this invariance of memory bandwidth from frequency has
significant ramifications for energy usage for bandwidth-limited algorithms:
absent the need for a high clock frequency to perform actual processing, the
CPU frequency can be lowered, thereby decreasing power consumption, while the
memory bandwidth stays constant. The consequences for energy usage are
illustrated in the heat maps shown in
Figures~\ref{fig:heat-e2s}~and~\ref{fig:heat-edp}.  The former illustrates the
required energy-to-solution to compute the Stream triad for a dataset size of
10\,GB on Sandy Bridge, Ivy Bridge, and Haswell microarchitectures. We observe
that the so-called race-to-idle is not very efficient for all three
microarchitectures. In terms of energy-to-solution adding more CPU speed makes
no sense as soon as main memory bandwidth is saturated. We find that for the
Sandy and Ivy Bridge microarchitectures, using half the number of available
cores at moderate clock frequencies provides the best energy-to-solution
result; for Haswell, using the lowest possible CPU frequency is viable.
Overall, for the Stream triad Haswell offers an improvement of 23\%
respectively 12\% over the Sandy and Ivy Bridge microarchitectures when it
comes to energy consumption. The improvement becomes even more pronounced when
taking the runtime into account: the \ac{EDP} metric weighs the consumed energy
by the total runtime to include time-constraints that are typically found in
HPC scenarios. For Sandy and Ivy Bridge we observe that the optimum EDP
solution requires high clock frequencies in order to lower the runtime;
however, increasing the clock frequency will result in higher energy usage,
thus increasing the other input factor of the \ac{EDP}. As we have shown
previously, on Haswell the sustained bandwidth is independent of CPU
frequency, which is why a very low frequency can be used. As a result, Haswell
outperforms the Sandy and Ivy Bridge microarchitectures by 55\% and 35\%
respectively in terms of \ac{EDP}.

\subsection{Cluster on Die}

As shown previously in Section~\ref{sec:hsw:package}, the CPU cores are
arranged around two rings with each ring having a dedicated memory controller.
In \ac{CoD} mode, the cores get equally separated into two memory domains.
This means that each core will only use a single memory controller.  To keep
latencies low, the general approach is to make a core access main memory
through the memory controller attached to its ring. However, with the number of
cores in the affinity domains being equal, the asymmetric core count on the two
physical rings makes exceptions necessary. In the design shown in
Fig.~\ref{fig:hsw-package} the 14 cores are divided into two affinity domains
of 7 cores each.  Using a simple load-benchmark together with
\texttt{likwid-perfctr} \cite{Treibig:2011:3} to access performance counters
and measure the number of memory accesses for each individual memory channel,
we find that cores 0--6 access main memory through the memory channels
associated with the memory controller on the left ring, and cores 7--13 those
associated with the memory controller on ring 1. Thus, only a core number 7 has
to take a detour across rings to access data from main memory. With \ac{CoD}
active the \ac{LLC} also becomes segmented. As each affinity domain contains
seven \ac{LLC} segments (2.5\,MB each), the total amount of \ac{LLC} for each
domain is 17.5\,MB---exactly half of the total amount of 35\,MB.

The \ac{CoD} mode is intended for highly optimized \ac{NUMA} codes and serves
two purposes: The first is to decrease latency by reducing the amount of
endpoints in the affinity domain. Instead of 14 LLC segments, data will be
distributed in only 7 segments inside each affinity domain, thereby decreasing
the mean hop count. Also, the requirement to pass through the two buffers
connecting the rings is eliminated for all but one \ac{LLC} segment.  The
second benefit of \ac{CoD} is to that bandwidth is increased by reducing the
probability of ring collisions that is implied by lowering the number of
participants from 14 to~7.

\section{ECM Model}

The \ac{ECM} model~\cite{Treibig:2009,hager:cpe,sthw15} is a simple resource
oriented analytical performance model focusing on the elementary resources
instruction throughput and data transfers. It can predict the runtime for a
steady-state execution of a loop body and can break down different runtime
contributions from execution and data transfers. The \ac{ECM} model is a
lightspeed model: It puts a optimal throughput assumption on instruction
execution and assumes that all data transfers are bandwidth limited. Any
hazards, dependencies and latency influences are neglected. Setting up the
model requires intimate knowledge about execution capabilities, data paths and
bandwidth values for the complete memory hierarchy. This involves sometimes
information beyond the vendor specification data sheet.

\subsection{Model input, construction, and assumptions}

The total estimated runtime is decomposed into execution time and data transfer
times. There are rules when contributions can overlap with each other. Times
are always in CPU core cycles. This is convenient as everything on a processor
happens in units of cycles and thus the model is independent of a specific
variant of the processor. Modern processors have multiple clock domains, cores,
caches and memory might have a different clock speed. For memory transfers the
time is converted to the standard bandwidth unit bytes/cycle. While different
clock domains make it more complicated to set up the model, the generic
formulation of the model supports it. The granularity of data transfers inside
the cache/memory hierarchy is that of cache lines (CL). As a consequence the
ECM model considers instructions equivalent to process one \ac{CL} length. Note
that a kernel might involve multiple data streams and therefore also multiple
\ac{CL}s.


The in-core execution and transfer times must be put together to arrive at a
prediction of single-thread execution time.  If $T_\mathrm{data}$ is the
transfer time, $T_\mathrm{OL}$ is the part of the core execution that overlaps
with the transfer time, and $T_\mathrm{nOL}$ is the part that does not, then

\begin{equation}
T_\mathrm{core} =  \max\left(T_\mathrm{nOL},T_\mathrm{OL}\right)\quad\mbox{and}\quad
T_\mathrm{ECM} =  \max(T_\mathrm{nOL}+T_\mathrm{data},T_\mathrm{OL})\label{eq:T}~.
\end{equation}

The model assumes that (i) core cycles in which loads are retired do not
overlap with any other data transfer in the memory hierarchy, but all other
in-core cycles (including pipeline bubbles) do, and (ii) the transfer times up
to the L1 cache are mutually non-overlapping.

A shorthand notation is used to summarize the relevant information about the
cycle times that comprise the model for a loop: We write the model as
\ecm{T_\mathrm{OL}}{T_\mathrm{nOL}}{T_\mathrm{L1L2}}{T_\mathrm{L2L3}}{T_\mathrm{L3Mem}}{},
where $T_\mathrm{nOL}$ and $T_\mathrm{OL}$ are as defined above, and the other
quantities are the data transfer times between adjacent memory hierarchy
levels. Cycle predictions for data sets fitting into any given memory level
can be calculated from this by adding up the appropriate contributions from
$T_\mathrm{data}$ and $T_\mathrm{nOL}$ and applying (\ref{eq:T}).  For
instance, if the ECM model reads \ecm{2}{4}{4}{4}{9}{cycles}, the prediction
for L2 cache will be $\max\left(2,4+4\right)\,cycles=8\,cycles$.  As a
shorthand notation for predictions we use a similar format but with
``$\epsep$'' as the delimiter. For the above example this would read as
$T_\mathrm{ECM}=\ecmp{4}{8}{12}{21}{cycles}$. Converting from time (cycles) to
performance is done by dividing the work $W$ (e.g., flops) by the runtime:
$P=W/T_\mathrm{ECM}$.  If $T_\mathrm{ECM}$ is given in clock cycles but the
desired unit of performance is F/s, we have to multiply by the clock speed.

\subsection{Chip-level bottleneck and saturation}

We assume that the single-core performance scales linearly until a bottleneck
is hit. On modern Intel processors the only bottleneck is the memory bandwidth,
which means that an upper performance limit is given by the Roof\/line
prediction for memory-bound execution: $P_\mathrm{BW} = I\cdot b_\mathrm S$,
where $I$ is  the computational intensity of the loop code.  The performance
scaling for $n$ cores is thus described by
$P(n)=\min\left(nP_\mathrm{ECM}^\mathrm{mem},I\cdot b_\mathrm S\right)$ if
$P_\mathrm{ECM}^\mathrm{mem}$ is the ECM model prediction for data in main memory.
The performance will saturate at
$n_\mathrm{S}=\left\lceil T_\mathrm{ECM}^\mathrm{mem}/T_\mathrm{L3Mem}\right\rceil$ cores.
\begin{equation}
n_\mathrm S = \left\lceil\frac{I\cdot b_\mathrm S}{P_\mathrm{ECM}^\mathrm{mem}}
	\right\rceil
 = \left\lceil\frac{T_\mathrm{ECM}^\mathrm{mem}}{T_\mathrm{L3Mem}}
	\right\rceil~.
\end{equation}

The ECM model~\cite{Treibig:2009,hager:cpe,sthw15} is an analytical performance
model for homogeneous code segments, mostly innermost loop kernels. It is
a light speed model and restricts the processor architecture to its elementary
resources: instruction execution and data transfers. While the model accounts
for hazards and dependencies in instruction execution it assumes perfect
streaming, neglecting latency or cache affects, on the data transfer side. In
this sense it is very similar to the roofline model \cite{roofline:2009}. In
contrast to the roofline model the ECM model takes into account all runtime
contributions from data transfers and uses a much more detailed view on
potential overlap among different runtime contributions. To set up the model
detailed knowledge about the code, the processor architecture and data volumes
and paths within the memory hierarchy. This process forces a developer
to learn more about his code and the processor architecture, which is an
important secondary benefit of the model compared to e.g. tool only approaches
where the outcome is a magic number without any insight or knowledge gain. As
a result the model provides detailed information about runtime contributions
and bottlenecks.

\subsection{Model setup}

The model operates on the level of processor work which are instructions and
transfered data volume. For this it is in most cases required to look at the
assembly level code. Within a cache hierarchy the smallest granularity of work
is one cacheline (usually 64b on X86 architectures). Work equivalent to one
cacheline length is also the granularity the ECM model operates on. The
primary time unit used in the model are processor core cycles. This is the
primary unit of time in a microarchitecture. To account for different clock
domains in modern processor designs other clock domains, e.g. DRAM or Uncore,
are converted into core cycles.

To set up the model the following steps must be performed:

\begin{enumerate}
    \item {\bf Determine the core cycles to execute the instructions which are required
        to process work equivalent to one cacheline length.}  In this context it is
        useful to look at work in terms of iterations on different levels. The
        first level is the operation level. Assume a memory copy is implemented
        in terms of double precision floating point assignments then the atomic
        operation is one double precision floating point copy. This is worth
        copying 8b of data. To update (or process) one cacheline as
        a consequence $64/8=8$ iterations on the operation level are required.
        Note that if we talk about one cacheline here it means to process work
        equivalent to one cacheline length. But of course multiple physical
        cachelines might be involved. For copy to process one cacheline
        results in reading from a source cacheline and storing to another
        destination cacheline. The number of iterations on the instruction
        level might be different though. If for example SIMD SSE instructions
        are used 16b can be copied with two instructions. Instead of 8b of one
        operation one instruction moves 16b. On the instruction level only
        4 iterations are needed to process one cacheline. The next level of
        iterations is the loop level. If a loop is unrolled multiple
        instruction iterations form one loop iteration. Lets assume the copy
        loop in our example is 4-times unrolled to update one cacheline length
        only one loop iteration is required. To wrap it up: To process one
        cacheline length requires one loop iteration which is equivalent to
        four instruction iterations which is equivalent to eight operation
        iterations. To determine the core cycles to throughput a sequence of
        instructions in a steady state a simple model of the instruction
        scheduler is required. The model does not limit the effort put into
        getting a sensible number for instruction throughput. For simple loop
        kernels this can be done by hand or in more complicated cases a simple
        simulator as the Intel IACA tool may be used. At this point it is
        assumed that all data is served from the L1 cache.
    \item {\bf Setup data paths and volumes to get the data to the L1 cache.}
        For streaming algorithms this step is rather simple. One needs to know
        about the store miss policy and overall cache architecture of the
        processor. If the store miss policy is write allocate additional
        cacheline transfers need to be accounted for. Intel processors have
        inclusive caches, data is always streamed through all cache levels. The
        store miss policy is write allocate up to the L1 cache. One must be
        careful as there exist special non-temporal store instructions for
        memory which do not trigger the write allocate. In contrast many
        competitors (AMD and IBM) use a write-through policy for the L1 cache.
        All data is initially loaded into L2 cache and the last level L3 cache
        is a victim cache. Only cachelines evicted from L2 are placed in L3.
        Things get more complex if data access is not pure streaming. This is
        the case for stencil codes which expose data reuse within the cache
        hierarchy. Sometimes data volumes are difficult to acquire. One
        solution is to validate data volumes with hardware performance counter
        measurements which allow to determine data volumes between different
        memory hierarchy levels.
    \item {\bf Setup the overall single core prediction by accounting for overlap.}
    \item {\bf Determine multicore scaling within a chip.}
\end{enumerate}



\section{Microbenchmarks}

The set of microbenchmarks used to verify the ECM model on the Haswell
microarchitecture is summarized in Table~\ref{tab:benchmarks}. In addition to
the loop body, the table lists the number of load and store streams---the
former being divided into explicit and \ac{RFO} streams. \ac{RFO}
refers to implicit loads that occur whenever a store miss in the L1 cache
triggers a write-allocate of the cache line required for the store. Also
included in the table are the predictions of the ECM model and the actually
measured runtimes in cycles per second along with a quantification of the
model's error. 

\begin{table*}
\renewcommand{\arraystretch}{1.2}      
\caption{Overview of microbenchmarks: Loop Body, Memory Streams, ECM prediction
and Measurement in c/CL, and Model Error.}
\label{tab:benchmarks}
\centering
\rowcolors{2}{gray!20}{white}
\resizebox{\linewidth}{!}{ 
\begin{tabular}{llccccc}
\hline
                    &                           & Load Streams      & Write     & ECM Prediction            & Measurement                      & Error \\
    Benchmark       & Description               & Explicit / RFO    & Streams   & L1/L2/L3/Mem              & L1/L2/L3/Mem                     & L1/L2/L3/Mem \\
\hline
    ddot            & \texttt{s+=A[i]*B[i]}     & 2 / 0             & 0         & \ecmp{2}{4}{8}{17.1}{}    & \ecmp{2.1}{4.7}{9.6}{19.4}{}     & \ecmp{5\%}{17\%}{20\%}{13\%}{} \\
    load            & \texttt{s+=A[i]}          & 1 / 0             & 0         & \ecmp{2}{2}{4}{8.5}{}     & \ecmp{2}{2.3}{5}{10.5}{}         & \ecmp{0\%}{15\%}{25\%}{23\%}{} \\
    store           & \texttt{A[i]=s}           & 0 / 1             & 1         & \ecmp{2}{5}{9}{21.5}{}    & \ecmp{2}{6}{8.2}{17.7}{}         & \ecmp{0\%}{20\%}{9\%}{19\%}{} \\
    update          & \texttt{A[i]=s*A[i]}      & 1 / 0             & 1         & \ecmp{2}{5}{9}{21.5}{}    & \ecmp{2.1}{6.5}{8.3}{17.6}{}     & \ecmp{5\%}{30\%}{8\%}{18\%}{} \\
    copy            & \texttt{A[i]=B[i]}        & 1 / 1             & 1         & \ecmp{2}{6}{12}{28.8}{}   & \ecmp{2.1}{8}{13}{27}{}          & \ecmp{5\%}{33\%}{8\%}{6\%}{} \\
    STREAM triad    & \texttt{A[i]=B[i]+s*C[i]} & 2 / 1             & 1         & \ecmp{3}{8}{16}{37.7}{}   & \ecmp{3.1}{10}{17.5}{37}{}       & \ecmp{3\%}{25\%}{9\%}{2\%}{} \\
    Sch\"onauer triad& \texttt{A[i]=B[i]+C[i]*D[i]} & 3 / 1         & 1         & \ecmp{4}{10}{20}{46.5}{}  & \ecmp{4.1}{11.9}{21.9}{46.8}{}   & \ecmp{3\%}{19\%}{9\%}{1\%}{} \\
\hline
\end{tabular}
} 
\end{table*}

The set of benchmarks contains a number of different streaming kernels, each
one offering a different combination of the different stream types to cover
different transfer scenarios in the cache hierarchy. In the following, we will
discuss and formulate the ECM model for each of the benchmarks. Note that the
sustained bandwidths used to derive the L3-memory cycles per \ac{CL} inputs are
that of a single memory domain---i.e. the seven cores comprising one memory
domain in \ac{CoD} mode---and not the sustained chip bandwidth.  We use the
\ac{CoD} mode, because it offers better performance than the non-\ac{CoD} mode.

\subsection{Dot Product and Load}

The dot product benchmark \textit{ddot} is a load-only benchmark that makes use
of the new FMA instructions introduced in the FMA3 ISA extension.  For this
benchmark $T_\mathrm{nOL}$ is two clock cycles, because the core has to load
two cache lines (\texttt{A} and \texttt{B}) from L1 to registers using four AVX
loads (which can be processed in two clock cycles, because each individual AVX
load can be retired in a single clock cycle and there are two load ports).
Processing the data from the cache lines using two AVX fused multiply-add
instructions only takes one clock cycle, because both issue ports~0~and~1
feature AVX FMA units. A total of two cache lines has to be transfered between
the adjacent cache levels. At 64\,B/c this means 2\,cy to transfer the CLs from
L2 to L1.  Transferring the CLs from L3 to L2 takes 4\,cy at 32\,B/c. The
empirically determined sustained (memory domain) bandwidth for the dot product
was 32.4\,GB/s.  At 2.3\,GHz, this corresponds to a bandwidth of about
4.5\,cy/CL or 9.1\,cy for two CLs. The ECM model input is thus
\ecm{1}{2}{2}{4}{9.1}{\mathrm{cycles}} and the corresponding prediction is
$T_\mathrm{ECM}=\ecmp{2}{4}{8}{17.1}{\mathrm{cycles}}$.

As the name suggest, the \textit{load} benchmark is a load-only benchmark as
well. However, here $T_\mathrm{nOL}$ and $T_\mathrm{OL}$ are interchanged:
while a single clock cycle suffices to load the elements from cache line
\texttt{A} into AVX registers, two cycles are required to process the data,
because there is only a single AVX add unit. Because only a single cache line
has to be transferred between adjacent cache levels, the time required is
exactly half of that needed for the \textit{ddot} benchmark.\footnote{The
sustained chip bandwidth is identical to that of the dot product
microbenchmark, resulting in the same memory bandwidth of 4.5\,c/CL.} The ECM
model input for the \textit{load} benchmark is
\ecm{2}{1}{1}{2}{4.5}{\mathrm{cycles}}. The model prediction is
$T_\mathrm{ECM}=\ecmp{2}{2}{4}{8.5}{\mathrm{cycles}}$.

\subsection{Store, Update, and Copy}

Using AVX instructions storing one cache line worth of constants for the
\textit{store} benchmark takes two clock cycles, because only one store unit is
available, resulting in $T_\mathrm{nOL} = 2$\,cy. As there are no other
instructions such as arithmetic operations, $T_\mathrm{nOL}$ is zero. When
counting cache line transfers along the cache hierarchy, we have to bear in
mind that a store-miss will trigger a write-allocate, thus resulting in two
cache line transfers for each cache line update: one to write-allocate the
cache line which data gets written to and one to evict the modified cache line
once the cache becomes full. Because evictions between from L1 to L2 cache take
place at a bandwidth of only 32\,B/c, this results in a transfer time of three
cycles to move cache lines between the L1 and L2 cache and a transfer time of 4
cycles for L2 and L3.  The sustained bandwidth for a benchmark involving
evictions is slightly worse than that of load-only kernels. In CPU cycles the
measured bandwidth of about 23.6\,GB/s corresponds to approximately 6.2\,cy/CL.
The resulting ECM input and prediction are
\ecm{0}{2}{3}{4}{12.5}{\mathrm{cycles}} respectively
\ecmp{2}{5}{9}{21.5}{\mathrm{cycles}}.

As far as the ECM model is concerned, the \textit{update} and \textit{store}
kernels behave very similar. The time required to perform a cache line update
is $T_\mathrm{nOL} = 2$\,cy as well, limited by store throughput. The two AVX
loads required to load the values to be updated can be performed in parallel to
the two store instructions. In addition, the stores are paired with the two AVX
multiplications required to update the values in the cache line, resulting in
$T_\mathrm{OL} = 2$\,cy.\footnote{Note that another pairing, such as e.g. one
store with two multiplications and one store with two loads is not possible due
to the limited number of full AGUs.} The number of cache line transfers is
identical to that of the \textit{store} kernel, the only difference being that
the cache line load is caused by explicit loads and not a write-allocate. With
a memory bandwidth almost identical to that of the \textit{store} kernel, the
time to transfer a cache line between L3 and memory again is approximately
6.2\,c/CL, yielding an ECM input of \ecm{2}{2}{3}{4}{12.5}{\mathrm{cycles}} and
a prediction that is identical to that of the \textit{store} kernel.

The \textit{copy} kernel has to perform two AVX loads and two AVX stores to
copy one cache line. In this scenario, again, the single store port is the
bottleneck, yielding $T_\textrm{nOL}=2$\,cycles. Instead of transferring two
cache lines, as was the case in the \textit{store} and \textit{update} kernels,
the \textit{copy} kernel has to transfer three cache lines between adjacent
cache levels: load \texttt{B}, write-allocate and evict \texttt{A}. Loading two
cache lines at 64\,B/c and evicting at 32\,B/c from and to L2 takes a total of
4\,cycles; transferring three cache lines at 32\,B/c between L2 and L3 requires
6\,cycles.  With a slightly higher sustained memory bandwidth of 26.3\,GB/s
than those of the \textit{store} and \textit{update} kernels due to the higher
load-to-store ratio of the \textit{copy} kernel the time to transfer one cache
line between main memory and the last-level cache is approximately 5.6\,cy.
This results in the following input for the ECM model
\ecm{0}{2}{4}{6}{16.8}{\mathrm{cycles}}, which in turn yields a prediction of
\ecmp{2}{6}{12}{28.8}{\mathrm{cycles}}.

\subsection{Stream and Sch\"onauer Triads}

For the \textit{Stream Triad}, the \ac{AGU}s prove to be the bottleneck: While
the core can potentially retire four micro-ops per cycle, it is impossible to
schedule two AVX loads (each corresponding to one micro-op) and an AVX store
(corresponding to \textit{two} micro-ops) which uses indexed addressing,
because there are only two \ac{AGU}s available supporting this addressing mode.
The resulting $T_\mathrm{nOL}$ thus is not 2 but 3\,cycles to issue four AVX
loads (two each for cache lines \texttt{B} and \texttt{C}) and two AVX stores
(two for cache line \texttt{A}).  The required arithmetic of two \ac{FMA}s can
be performed in a single cycle, because two \ac{AVX} \ac{FMA} units are
available.  Data traffic between adjacent cache levels is four cache lines:
load cache lines containing \texttt{B} and \texttt{C}, write-allocate and evict
the cache line containing \texttt{A}. The measured sustained bandwidth of
27.1\,GB/s corresponds to approximately 5.4\,cy/CL---or about 21.7\,cy for all
four cache lines.  The input parameters for the ECM model are thus
\ecm{1}{3}{5}{8}{21.7}{\mathrm{cycles}} leading to the follow prediction:
\ecmp{3}{8}{16}{37.7}{\mathrm{cycles}}.

The Sch\"onauer Triad involves the same arithmetic as the Stream Triad with an
additional operand having to be loaded into registers.  Again the
address-generation units prove to be bottleneck. Now, six AVX loads
(corresponding to cache lines \texttt{B}, \texttt{C}, and \texttt{D}) and two
AVX stores (cache line \texttt{A}) have to be performed; the total of these
eight instructions have to share two AGUs, resulting in a $T_\mathrm{nOL}$ of
4\,cycles. The two AVX fused multiply-add instructions can be performed in a
single cycle.  Data transfers between adjacent caches correspond to five cache
liens: \texttt{B}, \texttt{C}, and \texttt{D} require loading while cache line
\texttt{A} needs to write-allocated and evicted. For the L1 cache, this results
in a transfer time of 6 cycles (four to load four cache lines, two to evict one
cache line). The L2 cache transfer time is 10 cycles. The measured sustained
memory bandwidth of 27.8\,GB/s corresponds to about 5.3\,cy/CL or 26.5\,cy for
all five cache lines.  The resulting ECM input parameters are thus
\ecm{1}{4}{6}{10}{26.5}{\mathrm{cycles}} and the resulting prediction is
\ecmp{4}{10}{20}{46.5}{\mathrm{cycles}}.

\section{Experimental Testbed}

A standard two-socket server based on the Haswell-EP microarchitecture was
chosen for evaluating the kernels. The machine uses two-way SMT and has
fourteen moderately clocked (2.3\,GHz base frequency) cores per socket. Sixteen
vector registers are available for use with \ac{SSE}, \ac{AVX}, and \ac{AVX}2.
Using floating-point arithmetic, each core can execute two \ac{FMA}
instructions per cycle leading to a peak performance of 16~\ac{DP} or
32~\ac{SP} \ac{Flops} per cycle.  Memory bandwidth is provided by means of a
ccNUMA memory subsystem with four DDR4-2166 memory channels per socket.  In
order to achieve best performance during benchmarking \ac{CoD} was activated
and \ac{UFS} was disabled.  A summary of the machine configurations can be
found in Table~\ref{tab:testbed}.  

\begin{table}
\renewcommand{\arraystretch}{1.2}      
\caption{Test Machine Configuration.}
\label{tab:testbed}
\centering
\rowcolors{1}{gray!20}{white}
\resizebox{\linewidth}{!}{%
\begin{tabular}{lc}
\hline
Microarchitecture               & Haswell-EP \\
Model                           & Xeon E5-2695 v3 \\
Release Date                    & Q3 2014 \\
\hline
Nominal/Turbo Clock
Speed (Single-Core)             & 2.3\,GHz/3.3\,GHz  \\
Cores/Threads                   & 14/28   \\
Major ISA Extensions            & SSE, AVX, AVX2, FMA3 \\
\hline
L1/L2/L3 Cache                  & 14$\times$32\,kB/14$\times$256\,kB/35\,MB \\
Memory Configuration            & 4 channels DDR4-2166 \\
Theoretical Memory Bandwidth    & 69.3\,GB/s \\
\hline
\end{tabular}
} 
\end{table}

\section{Results}

The results presented in this section were obtained using hand-written assembly
code that was benchmarked using \texttt{likwid-perfctr} \cite{Treibig:2011:3}
to guarantee reproducibility as compilers tend to perform well-meant
optimizations (such as producing \ac{SSE} instead of of \ac{AVX} loads in order
to lower the probability for split cache line loads) that can end up being
counter-productive thus resulting in non-optimal code even for the most simple
of kernels.

\subsection{Load, Dot Product}

\begin{figure}[tb]
\includegraphics[width=\linewidth]{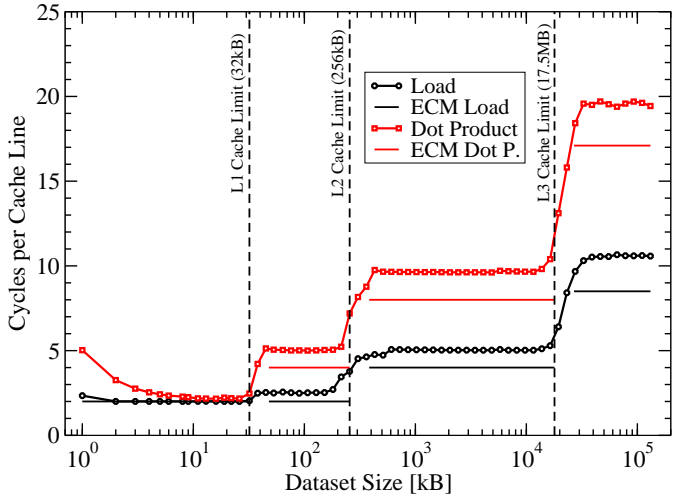}
\caption{ECM predictions and measurement results for load and dot product kernels.}
\label{fig:micro-loadddot}
\end{figure}

In Figure~\ref{fig:micro-loadddot} we illustrate ECM predictions and
measurement results for both the \textit{load} and \textit{dot product}
benchmarks. While the core execution time for both benchmarks is two clock
cycles just as predicted by the model, the \textit{dot product} performance is
slightly lower than predicted with data coming from the L2 cache. We found this
slightly worse than expected L2 cache performance to be a general problem with
Haswell.\footnote{In contrast to Haswell, both Sandy and Ivy Bridge's L2
bandwidth of 32\,B/c could be achieved in every benchmark \cite{sthw15}.}
In none of the cases the measured L2 performance could live up to the
advertised specs of 64\,B/c. However, the L2 performance is slightly better for
the \textit{load} benchmark. Here the performance in L2 is almost identical to
that with data residing in the L1 cache: this is because the cache line can
theoretically be transfered from L2 to L1 a single cycle at 64\,B/c, which is
exactly the amount of slack that is the difference between
$T_\mathrm{OL}=2$\,cy and $T_\mathrm{nOL}=1$\,cy. In practise, however, we
observe a small penalty of 0.3\,cy/CL.

As soon as the working set becomes too large for the core-local L2 cache, we
find that the ECM prediction becomes slightly off. An empirically determined
penalty for transferring data from off-core locations for kernels with a low
number of cycles per cache line was found to be one clock cycle per load stream
and cache-level, e.g. 2\,cy for the \textit{dot product} benchmark with data
residing in L3 and 4\,cy with data from main memory. This is most likely to be
attributed to additional latencies introduced when data is passing between
different clock domains (e.g. core, cbox, mbox) that can not entirely hidden
for kernels with a very low core cycle count.

\subsection{Store, Update, Copy}

\begin{figure}[]
\includegraphics[width=\linewidth]{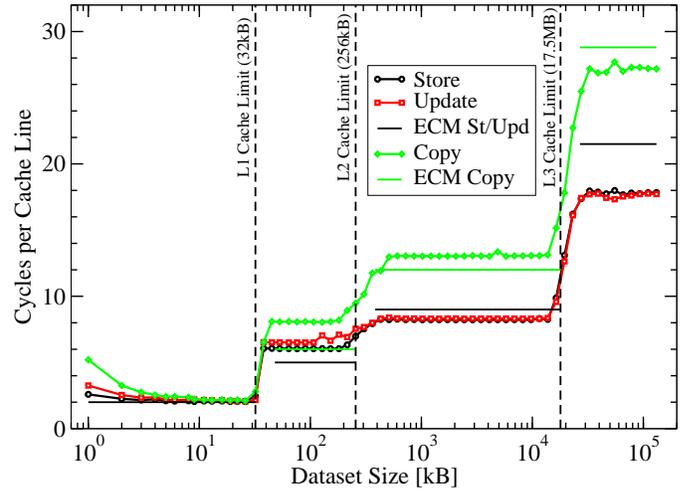}
\caption{ECM predictions and measurement results for store, update, and copy kernels.}
\label{fig:micro-stupco}
\end{figure}

In Figure~\ref{fig:micro-stupco} the ECM predictions and measurements for the
\textit{Store}, \textit{Update}, and \textit{Copy} kernels are shown. With data
coming from the L1 cache, the measurements for all three benchmarks matches the
model's prediction. As was the case previously, the measured performance is off
about one cycle per cache line loaded from L2 to L1 when data resides in the L2
cache: one cycle for the \textit{store} and \textit{update} benchmarks, and two
cycles for the \textit{copy} benchmark. As before, we attribute this to the
sustained L2 load bandwidth being lower than advertised.

Interestingly, the measured performance for the \textit{Store} and
\textit{Update} kernels in L2 is \textit{better} than the model prediction.  We
can rule out an undocumented optimization that avoids write-allocates when
rapidly overwriting cache lines in the L3, because the \textit{Store} kernel has
exactly the same performance as the \textit{Update} kernel, which has to load
the cache line in order to update the values contained in it. The \textit{Copy} 
kernel is about 1 cycle slower per cache line than predicted by the model.

For main memory, the measured result is significantly better than the model
prediction. This is caused by caches and several store buffers still holding
data to be evicted to main memory when the benchmark has completed.  Although
there exists a means to write-back all modified cache lines from caches to main
memory using the \texttt{wbinvd} instruction, the eviction will occur
asynchronously in the background, thereby making it impossible to measure the
exact time it takes to complete the benchmark.

\subsection{Stream Triad and Sch\"onauer Triad}
\label{sec:res:triads}

\begin{figure}[]
\includegraphics[width=\linewidth]{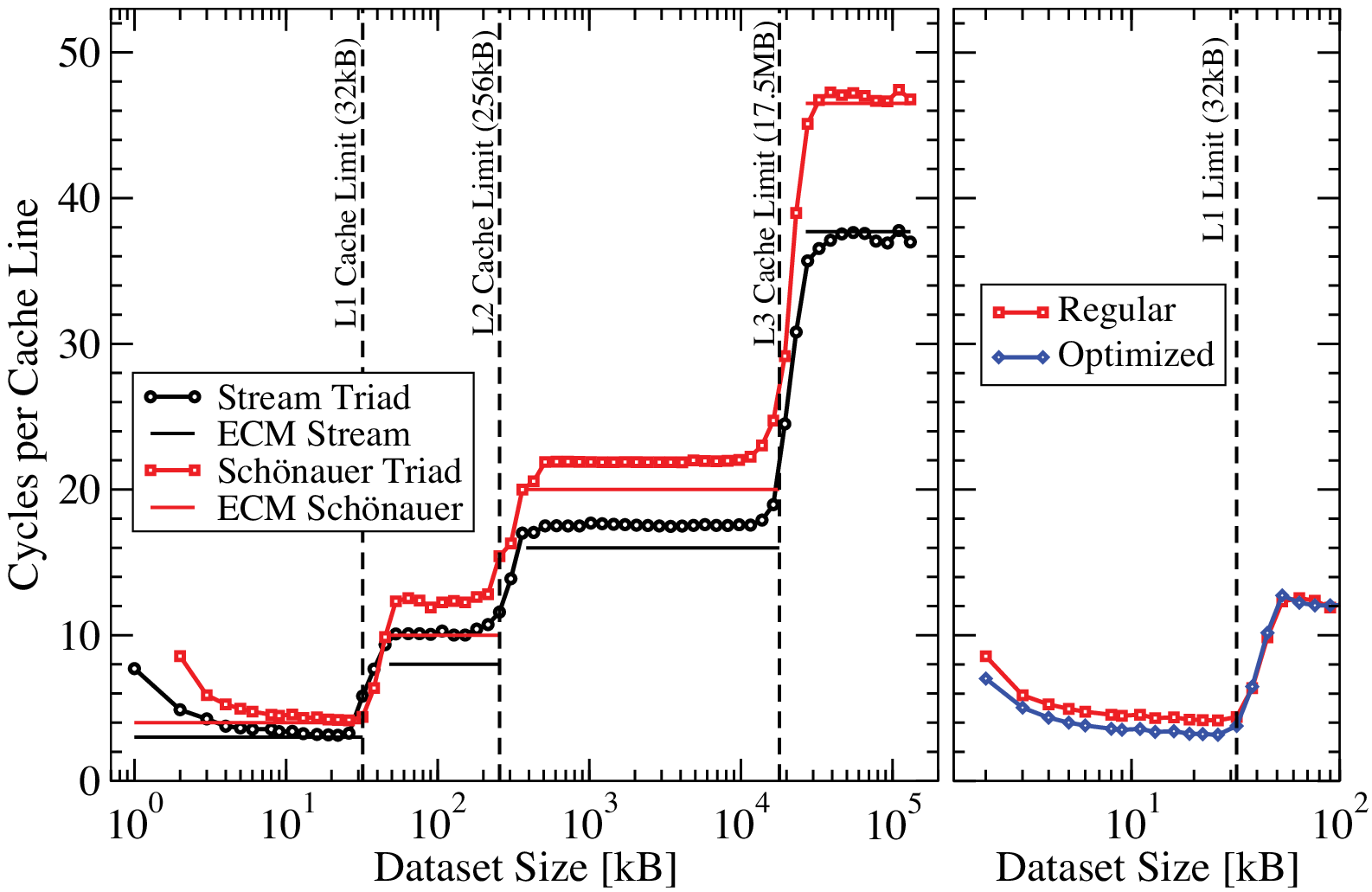}
\caption{ECM predictions and measurement results for Stream and Sch\"onauer
Triads (left) and comparison of naive and optimized Sch\"onauer Triad (right).}
\label{fig:micro-triads}
\end{figure}

In Figure~\ref{fig:micro-triads} we show the model predictions and actual
measurements for both the Stream and Sch\"onauer Triads. The measurement fits
the model's prediction with data in the L1 cache. As before, we observe the one
cycle penalty for each cache line that is loaded from the L2 cache, which
trickles down to the L3 cache as well. The measurement with data coming from
and going to main memory almost perfectly fits the model prediction.

In addition, Figure~\ref{fig:micro-triads} shows the measurement results for
the naive Sch\"onauer Triad as it is currently generated by compilers (e.g. the
Intel C Compiler 15.0.1) and an optimized version that makes use of all three
\ac{AGU}s, i.e. one that uses the newly introduced simple \ac{AGU} on port~7.
Typically, address calculations in loop-unrolled streaming kernels requires two
steps: scaling and offset computation. The scaling part involves multiplying
the loop counter with the size of the data type and adding it to a base address
(typically a pointer to the first element of the array) to compute the correct
byte address of $i$th array element; the offset part adds a fixed offset, e.g.
32\,B, to skip ahead the size of one vector register. Both \ac{AGU}s on ports~2
and~3 support this addressing mode called ``base plus index plus offset.'' The
problem with the simple \ac{AGU} is that it can not perform the indexing
operation but only offset computation.  However, it is possible to make use of
this \ac{AGU} by using one of the ``fast LEA'' units (which can perform
\textit{only} indexed and no offset addressing) to pre-compute an intermediary
address. This pre-computed address is fed to the simple \ac{AGU}, which can
then perform the still outstanding offset addition.  Using all three \ac{AGU}s,
it is possible to complete the eight addressing operations required for the
load/store operations in three instead of four cycles. The assembly code for
this optimized version is shown in Listing~\ref{src:schoenauer_lea}. Note that
due to lack of space we present only a two-way unrolled version of the kernel
instead of the eighy-way unrolled variant that was used for benchmarking.

\lstset{
        breaklines=true,
        language=C,
        basicstyle=\small\ttfamily,
        numbers=left,
        numberstyle=\tiny,
        frame=tb,
        columns=fullflexible,
        showstringspaces=false
}
\begin{lstlisting}[caption={Two-way unrolled, hand-optimized code for
Sch\"onauer Triad.},
    label=src:schoenauer_lea,
    float=htpb,
    captionpos=b,
    belowcaptionskip=4pt,
    ]
lea rbx, [r8+rax*8]
vmovapd ymm0, [rsi+rax*8]
vmovapd ymm1, [rsi+rax*8+0x20]
vmovapd ymm8, [rdx+rax*8]
vmovapd ymm9, [rdx+rax*8+0x20]
vfmadd231pd ymm0, ymm8, [rcx+rax*8]
vfmadd231pd ymm1, ymm9, [rcx+rax*8+0x20]
vmovapd [rbx], ymm0
vmovapd [rbx+0x20], ymm1
\end{lstlisting}

\subsection{Multi-Core Scaling}

\begin{figure}[tb]
\includegraphics[width=\linewidth]{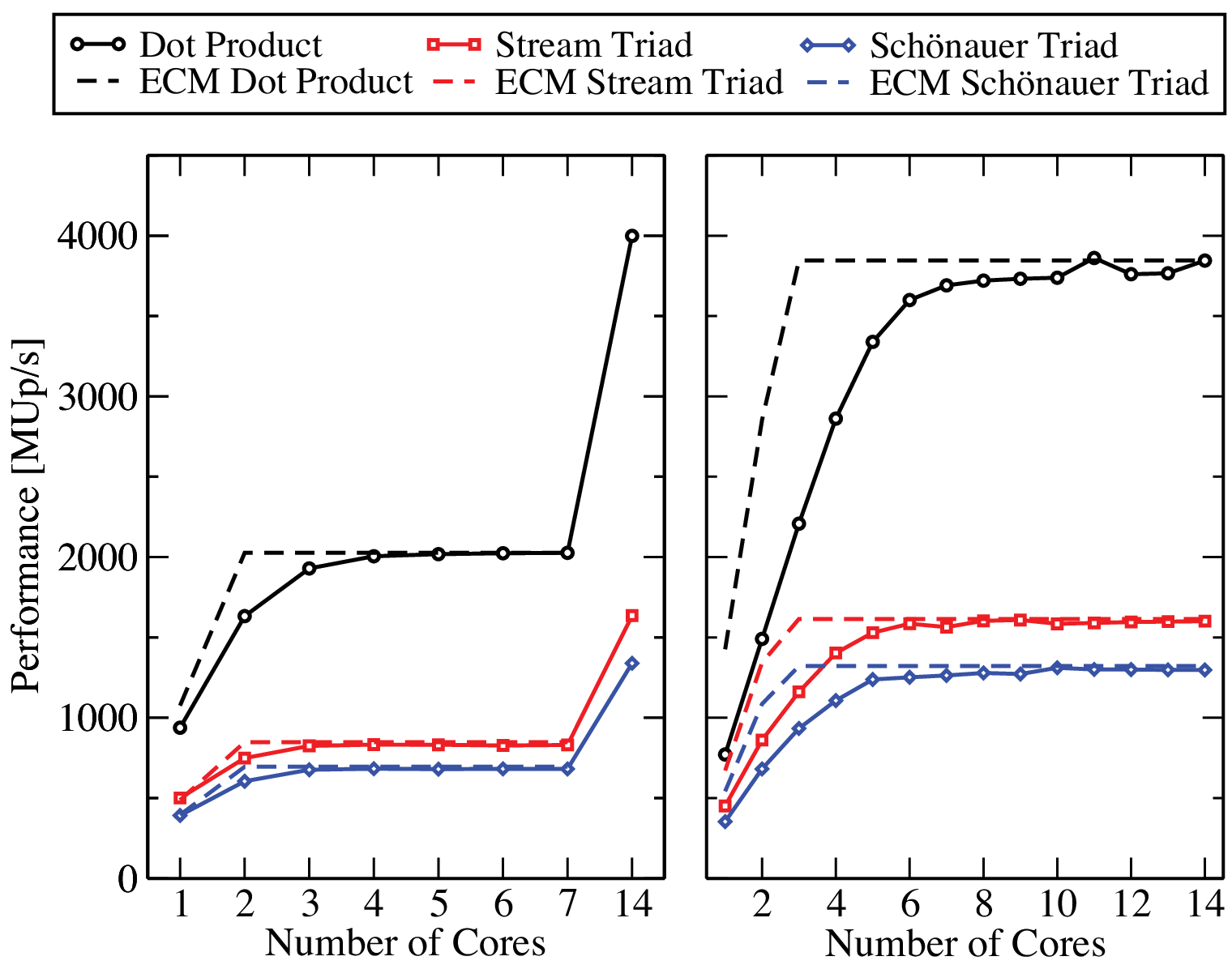}
\caption{Core-Scaling using \ac{CoD} mode (left) and non-\ac{CoD} mode (right).}
\label{fig:ecm-scaling}
\end{figure}

As discussed previously, when using the ECM to estimate multi-core performance,
single-core is scaled performance until a bottleneck is hit---which on Haswell
and other modern Intel CPUs is main memory bandwidth.
Figure~\ref{fig:ecm-scaling} shows ECM predictions along with actual
measurements for the \textit{dot product}, \textit{Stream Triad}, and
\textit{Sch\"onauer Triad} benchmarks using both \ac{CoD} and non-\ac{CoD}
modes.

The L3-memory transfer times for \ac{CoD} and non-\ac{CoD} mode have to be
based on the respective bandwidths of the mode. Transferring a cache line using
only one memory controller (in \ac{CoD} mode) takes more cycles than when using
both (non-\ac{CoD} mode).  In addition to scaling within the memory domain,
chip performance (fourteen cores) is also shown in \ac{CoD} mode. 

The measurements indicate that peak performance for both modes is nearly
identical, e.g. for the \textit{dot product} performance saturates slightly
below 4000\,MUp/s for non-\ac{CoD} mode while \ac{CoD} saturates slightly above
the 4000 mark. Although the plots indicate the bandwidth saturation point is
reached earlier in \ac{CoD} mode, this conclusion is deceiving. While it only
takes four cores to saturate the memory bandwidth of an memory domain, a single
domain is only using two memory controllers; thus, saturating chip bandwidth
requires $2\times4$ threads to saturate \textit{both} memory domains, the same
amount of cores it takes to achieve the sustained bandwidth in non-\ac{CoD}
mode.

\subsection{Non-Temporal Stores}

For streaming kernels and dataset sizes that do not fit into the \ac{LLC} it is
imperative to use non-temporal stores in order to achieve the best performance.
Not only is the total amount of data to be transfered from memory reduced by
getting rid of \ac{RFO} stream(s), but in addition, non-temporal stores do not
have to travel through the whole cache hierarchy and thus do not consume
valuable bandwidth. On Haswell, non-temporal stores are written from the L1
cache into core-private \ac{LFB}s, from which data goes directly into main
memory.

\begin{figure*}[!t]
\includegraphics[width=\linewidth]{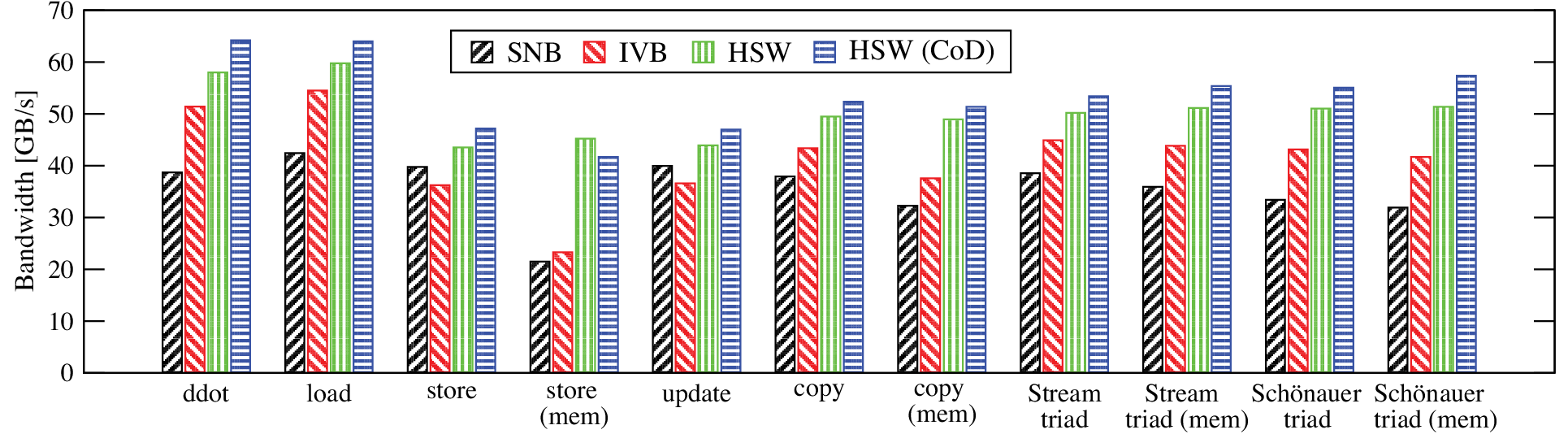}
\caption{Sustained Socket Bandwidth of Sandy Bridge (SNB), Ivy Bridge (IVB),
and Haswell both non-\ac{CoD} (HSW) and \ac{CoD} mode (HSW CoD)}
\label{fig:mem-efficiency-archs}
\end{figure*}

\begin{figure}[!b]
\includegraphics[width=\linewidth]{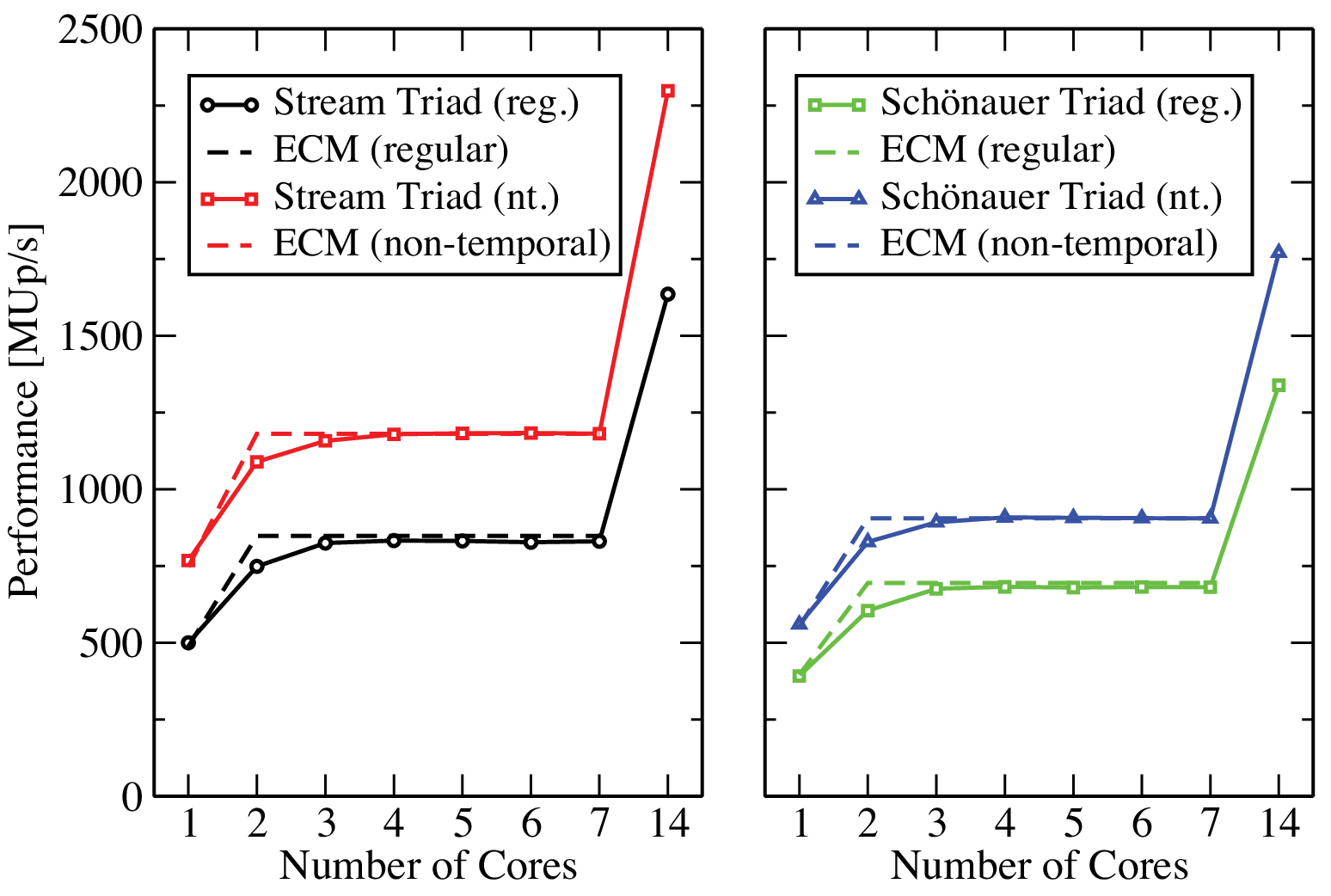}
\caption{Performance using regular vs. non-temporal stores for Stream (left)
and Sch\"onauer Triads (right).}
\label{fig:ecm-scaling-nt}
\end{figure}

Figure~\ref{fig:ecm-scaling-nt} shows the performance gain offered by
non-temporal stores. The left part shows the Stream Triad, which using regular
stores features two explicit load streams for arrays \texttt{B} and {C} plus a
store and an implicit \ac{RFO} stream for array \texttt{A}. Using the naive
roofline model, we would expect an increase of performance by a factor of
$1.33\times$, because employing non-temporal stores gets rid of the \ac{RFO}
stream, thereby reducing the number of streams from four to three. However, the
measured improvement in performance is higher: 1181\,MUp/s vs. 831\,MUp/s
($1.42\times$ faster) using a single affinity domain respectively 2298\,MUp/s
vs 1636\,MUp/s ($1.40\times$ faster) when using a full chip. This improvement
can not explained using a bandwidth-only model and requires accounting for
in-cache data transfers. Using non-temporal stores, the in-core execution time
stays the same. Instead of a L1-L2 transfer time of 5 cycles to load cache
lines containing \texttt{B} and \texttt{C} (2 cycles), write-allocating (1
cycle), and evicting (2 cycles) the cache line containing \texttt{A} the L1-L2
transfer time is now one just 4 cycles, because we don't have to write-allocate
\texttt{A}. The L2-L3 transfer time goes down from 8 cycles (load \texttt{B}
and \texttt{C}, write-allocate and evict \texttt{A}) to just 4 cycles (load
\texttt{B} and {C}). Also, cache line transfers to and from main memory go down
from four (load \texttt{B} and \texttt{C}, write-allocate and evict \texttt{A})
to three. At a sustained bandwidth of 28.3\,GB/s this corresponds to 5.2\,c/CL
or 15.6\,c/CL for three cache lines. The ECM input is thus
\ecm{1}{3}{4}{4}{15.6}{\mathrm{cycles}} leading to the follow prediction:
\ecmp{3}{7}{11}{26.6}{\mathrm{cycles}}. Comparing the 26.6\,c/CL with that of
the estimate of 37.7\,cy/CL when using regular stores (cf.
Table~\ref{tab:benchmarks}) we infer a speedup of exactly $1.42\times$ using
the ECM model.

We observe a similar behaviour for the Sch\"onauer Triad. Here, the roofline
model predicts an increase of performance by a factor of $1.25\times$ (four
streams instead of five). However, the measured performance using non-temporal
stores is 905\,GUp/s vs. 681\,GUp/s (factor $1.33\times$) using a single
affinity domain respectively 1770\,MUp/s vs. 1339\,MUp/s (factor $1.32\times$)
using a full chip. The ECM using non-temporal stores is constructed analogous
to the Stream Triad in the paragraph above. Three cache lines (\texttt{B},
\texttt{C}, and \texttt{D}) have to be transfered from L2 to L1; one cache line
\texttt{A} has to evicted from L1 to the \ac{LFB}s. Three cache lines
(\texttt{B}, \texttt{C}, and \texttt{D}) have to be transfered from L3 to L2.
Three cache lines (\texttt{B}, \texttt{C}, and \texttt{D}) have to be
transfered from memory to L3 and one cache line (\texttt{A}) has to be evicted
from the \ac{LFB}s to main memory. At a bandwidth of 29.0\,GB/s this
corresponds to approximately 5.1\,c per cache line or 20.3\,c for all four
cache lines.  The model input is thus \ecm{1}{4}{5}{6}{20.3}{\mathrm{cycles}},
yielding a prediction of \ecmp{4}{9}{15}{35.3}{\mathrm{cycles}}. Comparing
35.3\,c/CL to the estimate of 46.5\,cy/CL when using regular stores (cf.
Table~\ref{tab:benchmarks}) we infer a speedup of exactly $1.32\times$ using
the ECM model.

\subsection{Sustained Memory Bandwidth}

Apart from upgrading the memory from DDR~3 used in the previous Sandy and Ivy
Bridge microarchitectures to DDR~4 to increase the peak bandwidth, the
efficiency of the memory interface has been improved as well---especially with
regard to non-temporal stores. Figure~\ref{fig:mem-efficiency-archs} shows a
comparison of the sustained memory bandwidth achieved by the Haswell machine
(cf.  Table~\ref{tab:testbed}) and the predecessor microarchitectures Sandy and
Ivy Bridge. The Sandy and Ivy Bridge systems used for comparison are standard,
two-socket servers featuring Xeon E5-2680 (SNB) and Xeon E5-2690 v2 (IVY)
chips, with four memory channels per socket (DDR3-1600 in the Sandy Bridge and
DDR3-1866 in the Ivy Bridge node).

We observe that Haswell offers a higher bandwidth for all kernels, especially
when employing non-temporal stores. Also worth noting is that Haswell offers
improved bandwidth when using the \ac{CoD} mode for all but the store
benchmarks employing non-temporal stores.


\bibliographystyle{IEEEtran}

\end{document}